\documentclass [12pt,preprint]{aastex}
\usepackage{natbib}
\usepackage{epsfig}
\begin{document}

\voffset-0.5cm
\newcommand{\gsim}{\hbox{\rlap{$^>$}$_\sim$}}
\newcommand{\lsim}{\hbox{\rlap{$^<$}$_\sim$}}

\title{Correlated Spectral And Temporal Behaviour Of\\
Late-Time Afterglows Of Gamma Ray Bursts}

\author{Shlomo Dado\altaffilmark{1} and Arnon Dar\altaffilmark{1}}   

\altaffiltext{1}{Physics Department, Technion, Haifa 32000, Israel}   

\begin{abstract}

The cannonball (CB) model of gamma ray bursts (GRBs) predicts that the 
asymptotic behaviour of the spectral energy density of the X-ray 
afterglow of GRBs  is a power-law in time and in frequency where the 
difference between the temporal and spectral power-law indexes, 
$\alpha_X-\beta_X$, is restricted to the values 0, 1/2 and 1. Here we 
report the distributions of the values $\alpha_X$, $\beta_X$ and their 
difference for a sample of 315 Swift GRBs. This sample includes all 
Swift GRBs that were detected before August 1, 2012, whose X-ray afterglow 
extended well beyond 1 day and the estimated error in $\alpha_X-\beta_X$ 
was $\leq 0.25$. The values of $\alpha_X$ were extracted from the CB model 
fits to the entire light curves of their X-ray afterglow while the 
spectral index was extracted by the Swift team from the time integrated 
X-ray afterglow of these GRBs. We found that the distribution of 
the difference $\alpha_X-\beta_X$ for these 315 Swift GRBs has three narrow 
peaks around 0, 1/2 and 1 whose widths are consistent with being due to 
the measurement errors, in agreement with the CB model prediction.

\end{abstract} 


\maketitle

\section{Introduction}

A major breakthrough in the study of gamma ray bursts (GRBs) was the 
discovery of their X-ray afterglow by the Beppo-SAX satellite (Costa, E. 
et al. 1997), which led to their localization and consequently to the 
discovery of their longer wave-length afterglows (van Paradijs et al. 
1997, Frail \& Kulkarni~1997), which were predicted (Paczynski \& Roads 
1993, Katz 1994, Mezaros \& Rees 1997) by the fireball model of GRBs 
(Paczynski 1986, Goodman 1986). Until the launch of the Swift satellite, 
the fireball model was widely accepted as the correct model of GRBs and 
their afterglows (see, e.g., the reviews by Meszaros~2002, Zhang \& 
Meszaros~2004, Piran~2004). The rich data on GRBs and their afterglows 
obtained in recent years with the Swift and Fermi satellites complemented 
by data from ground-based rapid response telescopes and large follow-up 
telescopes, however, have challenged this prevailing view (see, e.g., 
Dar~2006 and the publications of Dado et al. cited in this paper and 
references therein; Margutti et al.~2012 and references therein).

In contrast, the cannonball model (CB) of GRBs (Dar \& De R\'ujula 2004 
and references therein) has been very successful in reproducing also the 
detailed light-curves of GRBs and their AGs that were measured with the 
Swift and Fermi satellites (see, e.g., Dado et al.~2009a,b). This success 
required adjustment of the free parameters of the CB model to fit 
the data, which could have made one wonder whether the good agreement 
between theory and observations was due to the flexibility of the model 
rather than to its validity.

The CB model, however, has been used also to predict universal properties 
of GRBs and their afterglows {\it before their discovery by observations}, 
which do not depend on free or adjustable parameters. For the prompt 
emission, these included the large linear polarization of the prompt gamma 
rays\footnote{Such evidence was reported by Coburn \& Boggs 2003 (see, 
however, Rutledge \& Fox~2004 and Wigger et al.~2004), Willis et al.~2005, 
McGlynn et al.~2007, Gotz et al.~2009 and Yonetoku et al. 2011.} (e.g., 
Shaviv \& Dar 1995), the rapid spectral softening that accompanied the 
fast decline of the prompt $\gamma$-ray emission (see, e.g., Dado et 
al.~2007 and references therein), and several correlations among GRB 
observables (see, e.g., Dar \& De R\'ujula~2000, Dado \& Dar~2012 and 
references therein) including the 'Amati relation' (Amati et al.~2002).

As for the afterglow emission, the CB model predicted the appearance
of an underlying supernova in the optical AG of 
relatively nearby long duration GRBs
(see, e.g., Dar 1999, Dado et al.~2002, 2003 and references therein), 
the canonical 
behaviour of X-ray light-curves - the transition to a plateau phase after 
a fast decline with a rapid softening of the prompt emission spectral 
energy density (SED) that
bends/breaks smoothly into an asymptotic power-law decline (e.g., Dado et 
al.~2002, Dar \& De R\'ujula~2004) - before it was discovered with Swift 
(Nousek et al.~2006), and the 
chromatic behaviour of the broad-band (XUVONIR) afterglow 
at early time (e.g., Covino et 
al.~2006) that becomes  achromatic well after the smooth 
bend/break time with roughly a universal spectral index $\beta_{UVONIR}\approx 
\beta_X\approx 1.1$\footnote{This behaviour was predicted only for 
late-time non-flaring afterglows.}. 

In this paper, we have used the CB model to derive another universal 
feature of the late-time broad band (XUVONIR) afterglow of GRBs, which 
does not involve adjustable parameters. Namely, we show that in the CB 
model the observed pectral energy density of the {\it late-time} afterglow 
declines with time and frequency, like $F_\nu(t)\sim t^{-\alpha_\nu}\, 
\nu^{-\beta_\nu}$ where the difference between the temporal and spectral 
power-law indexes can have one of the values $\alpha_\nu-\beta_\nu$=0, 1/2 
or 1. This closure relation then was tested using the 0.3-10 keV 
light-curves of the X-ray afterglow of 315 Swift GRBs, which were measured 
accurately enough well beyond 1 day with the Swift X-ray telescope (XRT).  
The CB model best fits to the entire available data of their X-ray 
afterglows were used to extract the values of their late-time temporal 
index $\alpha_X$, while the values of their spectral index $\beta_X$ were 
those inferred by the Swift team from the spectrum of the time integrated 
X-ray afterglow.

\section{The synchrotron radiation afterglow at late time}   
 
In the CB model (see, e.g., Dar \& De R\'ujula 2004, Dado et al.~2002, 
2009a and references therein) GRBs and their afterglows are produced by 
the interaction of bipolar jets of highly relativistic plasmoids (CBs) of 
ordinary matter with the radiation and matter along their trajectory 
(Shaviv \& Dar 1995, Dar 1998). Such jetted CBs are presumably ejected in 
accretion episodes on the newly formed compact stellar object in 
core-collapse supernova (SN) explosions (Dar et al.~1992, Dar \& 
Plaga~1999, Dar \& De R\'ujula~2000), in the merger of compact objects in 
close binary systems (Goodman et al.~1987, Shaviv \& Dar 1995) and in mass 
accretion episodes in microquasars and phase transitions in compact stars 
(Dar~1998, Dado et al. 2009b).

In the CB model,
the circumburst medium in front of a highly relativistic CB is completely 
ionized by the radiation from the CB. In the CB's  rest frame,
the ions of the medium that are continuously impinging on the CB
generate within it a turbulent magnetic field, whose energy density  is 
assumed to
be in approximate equipartition with that of the  impinging particles, 
$B\approx \sqrt{4\,\pi\, n\, m_p\, c^2}\, \gamma$ (e.g., Dado et al. 
2002),
where $n$ is the external baryon density and $m_p$ is the
proton mass.
The electrons that enter the CB with a relative Lorentz factor $\gamma'_e(t)=\gamma(t)$ 
in the CB's rest frame  are Fermi accelerated there 
to a smoothly broken power-law distribution in $\gamma'_e$ with a break around
$\gamma'_e(t)=\gamma(t)$  and a power-law index $p_e\sim 2.1 $ well above it.
These electrons cool rapidly by emission of synchrotron radiation (SR).
This  SR is
isotropic in the CB's rest frame and has a smoothly broken power-law
with  a characteristic break frequency $\nu'_b(t)$, which
is the typical synchrotron frequency radiated by the interstellar medium 
(ISM) electrons
that enter  the CB at time $t$ with a  Lorentz factor $\gamma(t)$.
In the observer frame, the emitted photons are beamed
into a narrow cone along the CB's direction of motion
by its highly relativistic bulk motion, their arrival times
are aberrated and their energies are boosted by
its  bulk motion Doppler factor $\delta$ and redshifted by the cosmic
expansion during their  travel time to the observer. In particular,
in the observer frame (see, e.g. Eq.~(25) in Dado et al.~2009a),
\begin{equation}
\nu_b(t)=\delta(t)\,\nu'_b(t)/(1+z)\propto n^{1/2}\,[\gamma(t)]^3
\delta(t)\,, 
\label{nub}
\end{equation}
where $\delta\! =\! 1/\gamma\, (1\!-\!\beta\, cos\theta)$ is
its Doppler factor with $\theta$ being the angle between the line of sight
to the CB and its direction of motion. For $\gamma^2 \gg 1$ and $\theta^2
\ll 1$, $\delta \approx 2\, \gamma/(1\!+\!\gamma^2\, \theta^2)$ to an
excellent approximation. The spectral energy density of the {\it unabsorbed} 
SR  afterglow has the form (see, e.g., Eq.~(26) in Dado et al.~2009a),
\begin{equation}
F_{\nu} \propto n^{(\beta_\nu+1)/2}\,[\gamma(t)]^{3\,\beta_\nu-1}\, 
[\delta(t)]^{\beta_\nu+3}\, \nu^{-\beta_\nu}\, ,
\label{Fnu}
\end{equation}
where the spectral index $\beta_\nu$ of the emitted radiation 
is related to the power-law index $p_e$ of the Fermi accelerated electrons
through  $\beta_\nu=p_e/2$ for $\nu > \nu_b$ and 
$\beta_\nu=(p_e-1)/2$ for $\nu < \nu_b$. 
The photon spectral index of the unabsorbed X-ray afterglow 
and the photon spectral index of the emitted radiation 
is given by $\Gamma_\nu=\beta_\nu+1$.  

The intercepted ISM particles that are swept into the CB decelerates its 
motion. For 
a CB of a baryon number $N_{_B}$, a radius $R$ and an initial Lorentz 
factor $\gamma_0\gg1$, which propagates in an ISM of a constant 
density 
$n$, relativistic energy-momentum conservation for plastic collisions
yields the deceleration law 
(Dado et al. 2009b and references therein):
\begin{equation}
\gamma(t) = {\gamma_0\over [\sqrt{(1+\theta^2\,\gamma_0^2)^2 +t/t_0}
          - \theta^2\,\gamma_0^2]^{1/2}}\,,
\label{goft}
\end{equation}
where $\gamma_0=\gamma(0)$ and $t_0={(1\!+\!z)\, N_{_{\rm B}}/ 8\,c\, 
n\,\pi\, R^2\,\gamma_0^3}\,.$

As can be seen from Eqs.~(2) and (3),
for a constant-density ISM,
the shape of the light-curve of the  AG depends on only  three parameters: the  product 
$\gamma_0\, \theta$,
the deceleration time scale $t_0$ and the spectral index $\beta(t)$.
As long as  $t\lsim t_b=(1+\theta^2\,\gamma_0^2)^2\, t_0$, 
$\gamma(t)$ and consequently $\delta(t)$, $\nu_b(t)$ and $\beta(t)$
change rather slowly with $t$ and generate the plateau phase of 
$F_\nu(t)$. For $t\gg t_b$, Eq.~(3) yields  
$\gamma(t)\propto t^{-1/4}$. Thus, for late time ($t\gg t_b$),
$[\gamma(t)\theta]^2$ becomes $\ll 1$, $\delta\approx 2\,\gamma(t)$  
and $\nu_b$ decreases to well below the UVONIR bands ($\beta\approx 
\beta_X\approx 
1.1$). Consequently,  Eq.~(2), yields 
a universal  achromatic power-law behaviour of the late-time ($t\gg t_b$)
broadband XUVONIR AG,   
\begin{equation}
F_\nu(t)\propto [\gamma(t)]^{(4\,\beta_X+2)}\,\nu^{-\beta_X}
        \propto  t^{-(\beta_X+1/2)}\, \nu^{-\beta_X}\,,
\label{Fnulate}
\end{equation}
where  $\alpha-\beta=\alpha_X-\beta_X=1/2$.
Such a canonical behviour of the X-ray afterglow of GRBs is
demonstrated in Figure 1 where the 0.3-10 keV X-ray light-curve of GRB 
060729 that
was measured with the Swift XRT is plotted together with its CB model best 
fit light-curve.

Often the break of the X-ray afterglow of very luminous GRBs is hidden 
under the prompt GRB emission or its fast decline phase (this happens 
because $\gamma_0$ is rather large and $[\gamma_0\, \theta]^2\ll 1$ 
yielding a rather small $t_b$). In such GRBs the "late-time"  power-law 
decline of the X-ray AG begins right after the fast decline of the prompt 
emission with the canonical late-time temporal index 
$\alpha_X=\beta_X+1/2$ (Dado et al. 2008). This is demonstrated in Figure 
2 where the light-curve of the X-ray afterglow of GRB 061007 that was 
measured with the Swift XRT is plotted together with its CB-model best fit 
light-curve.

Consider now a CB that encounters a  wind-like density profile 
$n\sim n_0\, (r_0/r)^2$ at $r=r_0$. For simplicity, let us set the 
encounter 
time to be $t=0$ in the observer frame. 
As long as the CB has not yet swept in a relativistic mass
comparable to its mass, $\gamma$ and $\delta$ change rather slowly
as a function of time and stay put at their values at $t=0$.  Hence,  
$r-r_0\approx\gamma(0)\, \delta(0)\, c\, t/(1+z)\propto t$.
Consequently, as long as the CB has not swept in 
a relativistic mass comparable to its rest mass,
Eqs.~(2) and (1) yield  for a  wind-like density profile 
\begin{equation}
F_{\nu} \propto  n^{(1+\beta_X)/2}\,\nu^{-\beta_X} 
        \propto t^{-(1+\beta_X)}\, \nu^{-\beta_X}\\. 
\label{Fnuwind}
\end{equation}
Consequently, $\alpha_X=\Gamma_X$, i.e.,  $\alpha_X-\beta_X$=1.  
(Dado et al. 2009a). 
A GRB produced near the center of an elliptical galaxy may
have an X-ray afterglow with such a late-time behaviour. This is 
demonstrated in Figure 4
where we plotted the light-curve of the X-ray afterglow of 
the short GRB 060414 and its CB model best fit light-curve.

In the CB model, the prompt emission pulses are produced by different CBs 
which, for simplicity, are assumed to be emitted in the same direction. 
The time resolution during the afterglow phase is usually much longer than 
the time separation between the prompt GRB pulses, i.e., the separation 
between the CBs' emission times, which, coupled with the initial fast 
expansion of the CBs and their merger due to the deceleration of the 
leading one in the ISM, allow us to consider them as a single blob during 
the afterglow phase (see e.g. Dar \& De R\'ujula 2004), which we have so 
far done.

However, if the ejecta has a "shot gun" distribution, i.e., consist of 
many ($N_{CB}\gg 1$)  CBs, 
which are spread within a small angle angle $\theta_e$ 
($ 1/\gamma(0)\ll\theta_e\ll 1$),
then one must 
integrate their contribution taking into account their different viewing 
angles. In the limit of a uniform angular distribution of CBs within an 
angle $\theta_e$ (namely, an angular density $N_{CB}/\pi\, \theta_e^2$) 
and neglecting the spread in arrival times from different CBs 
at late time, the simple 
integration of 
Eq.~(1) over $\theta$ for an on-axis observer yields the {\it late-time} 
SED,
\begin{equation}
F_\nu(t)\propto [\gamma(t)]^{4\,\beta_X}\,\nu^{-\beta_X}
        \propto  t^{-\beta_X}\, \nu^{-\beta_X}
\label{Fnusg}
\end{equation}
i.e., $\alpha_X=\beta_X$ and hence, $\alpha-\beta=\alpha_X-\beta_X=0$
and  $\nu_b(t)\sim t^{-\alpha_X/\beta_X}\rightarrow t^{-1}$ at late time.
These results are valid also for observers slightly off-axis
because the integration over viewing angle is dominated by the 
contribution from CBs moving near the line of sight.  
In Figure 5 we have plotted the 0.3-10 keV X-ray afterglow of
GRB 110808A that was retrieved from  the Swift/XRT light-curve repository 
(Evans et al. 2007, 2009) and  its CB-model best
fit light-curve  assuming a shot-gun configuration of CBs.

All together, the CB model predicts that the late-time temporal
decay of the GRB afterglow is well approximated by  power-law with a 
power-law index $\alpha_\nu$ that 
is simplly related to the spectral index $\beta_\nu$ of the late-time 
afterglow. In particular it predicts that
error-free measurements of the 
late-time $\alpha_X-\beta_X$, must yield a 
triple-peak distribution with narrow peaks around 0, 1/2 and 1 provided 
that the circumburst environments are well represented by the assumed 
constant density ISM or a wind-like density profile.

Note, however, that while the CB model predicts a triple peak distribution 
of $\alpha-\beta$ at late time, it predicts a single-peak distribution of 
the late-time $\beta$ around 1.083 and a late-time power-law decay of 
$\nu_b$ with a temporal index that is peaked around $-1.083$ (e.g., Dar 
and De R\'ujula 2004, 2007).

\section{Comparison with observations}
\noindent

Reliable and accurate values of the temporal and spectral power-law 
indexes $\alpha_X$ and $\beta_X$ of the late-time X-ray afterglow measured 
with the Swif XRT, which are reported in the
Swift/XRT light-curve repository (Evans et al. 2007, 2009), often are 
difficult to obtain because of one or more of the following reasons:\\
(a) The XRT measurements often run out of statistics or approach the 
background level before $\alpha_X$ reaches its late-time constant value.\\
(b) Temporal gaps in the late-time data.\\
(c) Flares that are superimposed on the presumably smooth afterglow 
light-curves.\\
(d) A too simplistic model of extragalactic absorption
restricted to the host galaxy whose chemical 
composition is taken to be solar and independent of redshift, that is used 
to infer $\Gamma_X$. Moreover, GRBs without known 
redshift are treated as GRBs at redshift z = 0. 

Before the launch of the Swift satellite, the knowledge of the X-ray 
afterglow of GRBs was incomplete and its inferred behaviour from limited 
and patchy data appeared to agree with the fireball model predictions 
(Meszaros and Rees 1997; Sari et al. 1998). After the launch of the Swift 
satellite it became rather clear that the observed properties and the 
diversity of the measured X-ray-afterglows, both at early and late times, 
challenge the fireball model paradigm (see, e.g., Dar 2006 and the other
publications of Dado et al. cited in the bibliography, and the references 
therein). This led to empirical parametrizations of AG light-curves such 
as smoothly broken power-laws (e.g., Beuermann et al. 1999) exponential to 
power-law functions (e.g., Willingale et al. 2007), and smoothly broken 
segmented power-laws (e.g., Margutti et al. 2012), which have never been 
derived from an underlying physical model. In most cases it was possible 
to fit reasonably well the observational data by introducing sufficient 
number of free adjustable parameters. However, because of difficulties 
(a)-(d) and the scarcity of late-time data it was not clear whether the 
extrapolation of these fits to late times yields reliable values of the 
late-time (asymptotic) $\alpha_X$.

In contrast, the CB model light-curves summarized in Eqs. (1)-(2), which 
were derived in fair approximations from the underlying assumptions of the 
CB model, were shown to describe very well the observed X-ray light-curves 
of the X-ray AGs of a large sample (over 150) of Swift GRBs with known 
redshift, including those with superimposed early-time and late- time 
flares (see, e.g., Dado et al. 2009a,b; Dado and Dar 2010 and references 
therein). In particular, their underlying smooth component was fitted very 
well by Eqs. (1)-(2) only with three adjustable parameters, $\gamma_0\, 
\theta$, $t_b$ and $p_e=2\, \beta_X$ and the properly chosen circumburst 
environment (a constant density or a wind-like density profile). This is 
demonstrated in Figures 1-6.

Thus, in order to avoid difficulties (a)-(d) and empirical 
parametrizations of the X-ray AG, we have extracted the late-time temporal 
slope $\alpha_X$ from the asymptotic power-law behaviour as given by Eqs.
(4)-(6) of the cannonball model fits to the entire X-ray light-curve
Moreover, since the late-time behaviour of the X-ray AG is independent of 
redshift, we have extended our fits to all the 315 X-ray light-curves that 
were reported in the Swift/XRT light-curve repository (Evans et al. 2007, 
2009) before August 1, 2012, with or without known redshift, whose 
measured 0.3-10 keV X-ray AG extended beyond 1 day and the reported error 
in their inferred spectral index is $\leq 0.25$. The typical error in the 
values of $\alpha_X$ obtained from the CB model fits to the entire 0.3-10 
keV light-curve of the X-ray AG of the Swift GRBs was much smaller than 
the 
typical error in the inferred value of $\Gamma_X$ that was reported in the 
Swift/XRT light-curve repository (Evans et al. 2007,2009).

We caution that the values of the late-time $\alpha_X$ that were 
obtained from the CB model best fit light-curves of the entire X-ray 
afterglow may differ
significantly from those obtained from empirical parametrizations of the
light-curves and arbitrary selection of time intervals. 
For instance, the CB model best fit light-curve
of the smooth 
0.3-10 keV light-curve of the  X-ray afterglow  of GRB 090618 that is 
shown in Figure 6,
yielded $\gamma_0\theta=0.707$, $t_b=2953$ s and $p_e=2.084$ i.e., a
late-time $\alpha_X=1.584\pm0.03$ ($\chi^2/dof=1.06$ for dof=1301) while a 
smoothly broken power-law (Beuermann et al. 1999) fit yields
$\alpha_1=0.79\pm 0.01$, a break time $t_b=0.50\pm 0.11$ d and a  
late-time $\alpha_X=\alpha_2=1.74\pm 0.04$ ($\chi^2/dof=1.17$).
The best fit spectral index of the unabsorbed 0.3-10 keV spectrum 
of the afterglow ($t > 250$ s) is $\beta_X=1.04\pm 0.04$. The CB model
prediction 
$\alpha_X=\beta_X+1/2=1.54\pm 0.04$  is consistent 
with the value  
$\alpha_X=1.584$ inferred from the CB model best fit light-curve of the
0.3-10 keV X-ray AG of GRB 090618 reported in the Swift/XRT light-curve 
repository (Evans et al.
2007, 2009), but is inconsistent with the value inferred from the 
smoothly broken power-law parametrization of the AG.

In Figure 7 we have plotted the values of $\Gamma_X$ of the 
time-integrated 0.3-10 keV X-ray AG of the 315 Swift GRBs, which were 
retrieved from the Swift/XRT light-curve repository. Their distribution is 
shown in Figure 8. As can be seen from Figures 7 and 8, their distribution 
peaks around the canonical value $\Gamma_X=25/12$ of the cannonball model, 
which is represented by the horizontal and vertical lines, respectively, 
in Figures 7 and 8.

In Figure 9 we have plotted the values of late-time slope $\alpha_X$ of 
the X-ray AG of the 315 GRBs plotted in Figures 7 and 8 that were obtained 
from the CB model best fit light-curves of the 0.3-10 keV 
X-ray AGs observed with the Swift XRT. These slopes seem to cluster around 
the canonical values 2.1, 1.6 and 1 of the CB model as is indicated by 
their distribution, which is plotted in Figure 10. This clustering is much 
more evident in Figures 11 and 12 where the values of $\alpha_X-\beta_X$ 
and their distribution are shown for the 315 GRBs plotted in Figure 7. The 
horizontal and vertical lines in Figures 11 and 12, respectively, 
represent the CB model expectation $\alpha_X-\beta_X=0$, 1/2, or 1 for 
error-free measurements.

\section{Conclusion}:
The spectral energy density of the X-ray afterglow of 315 ordinary GRBs 
detected before August 1, 2012, which have a late-time X-ray AG that was 
well measured with the Swift/XRT, is well described by a simple power-law 
in time and frequency, $F_\nu(t)\sim t^{-\alpha}\, \nu^{-\beta}$. Their 
X-ray afterglows are well described by the CB model (e.g., Dado et 
al.~2009a). The distribution of the difference between the late-time 
temporal and spectral power-law indexes $\alpha_X-\beta_X$ extracted from 
their CB model fits has a triple peak structure with peaks around 0, 1/2 
and 1 of full widths at half maximum consistent with the measurement 
errors while the distribution of $\beta_X$ has only a single peak around 
1.08. This behaviour is in good agreement with the CB-model prediction 
that the difference between the late-time temporal and spectral power-law 
indexes of GRBs is restricted to the values 0, 1/2 or 1, while such a 
behaviour challenges alternative GRB models.

{\bf Acknowledgement:} {The authors would like to thank an anonymous 
referee for useful comments and suggestions.}

\begin{deluxetable}{llllllll}
\tablewidth{0pt}
\tablecaption{The best fit parameters of the CB model description of 
the light-curves of the 0.3-10 keV X-ray afterglow of   
a representative sample of GRBs measured with Swift/XRT and 
shown in Figs 1-6. Listed also are their  
asymptotic temporal index $\alpha_X$ obtained from the fits
and their  photon spectral index $\Gamma_X$ measured 
with the Swit X-ray telescope (XRT).}
\tablehead{
\colhead{GRB} &\colhead{$t_i$[s]} & \colhead{$\gamma_0\theta$} &  
\colhead{$ t_{break}$[s]} & \colhead{$p_e$} &  \colhead{$\chi^2/dof $} & 
\colhead{$\alpha_X$}, & \colhead{$\Gamma_X$}\\}  
\startdata
060729  & 310 & 1.723 & 32876 & 2.096 & 1.39 &  1.548& $2.04 \pm  0.04$ \\
061007  &  40 & 0.908 & 135   & 2.052 & 1.00 &  1.526& $2.01 \pm  0.10$ \\
120422A & 276 & 1.203 & 84256 & 1.958 & 0.69 &  1.458& $2.01 \pm  0.25$ \\
060614  &     &       &       & 1.98  & 1.22 &  1.98 & $1.90 \pm  0.09$ \\
110808A & 100 &Shotgun & 17692 & 2.195 & 1.04 & 1.09 & $2.34 \pm  0.22$ \\
090618  & 310 & 0.708 & 2953  & 2.083 & 1.06 &  1.54 &  $2.00\pm  0.10$ \\
             
\enddata     
\end{deluxetable}

\newpage
\begin{figure}[]
\centering
\epsfig{file=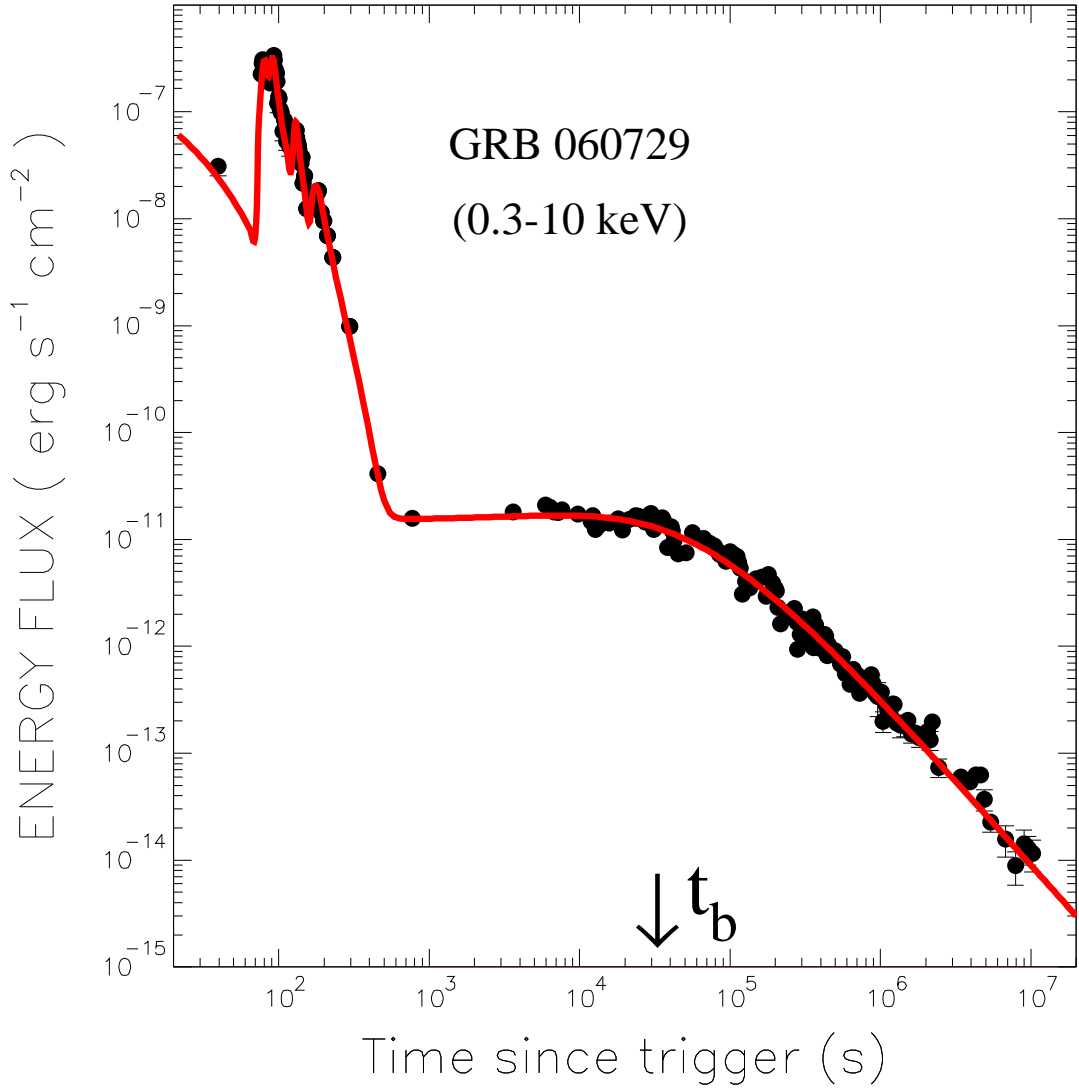,width=16.0cm,height=16.0cm}
\caption{
The X-ray light-curve of GRB 060729  and its CB model description
(Dado et al. 2009a) assuming  a constant density ISM and the late-time
value $\beta_X=1.02\pm 0.04$ reported in the Swift/XRT light-curve
repository (Evans et al. 2007,2009).}
\label{fig1}
\end{figure}

\begin{figure}[]
\centering
\epsfig{file=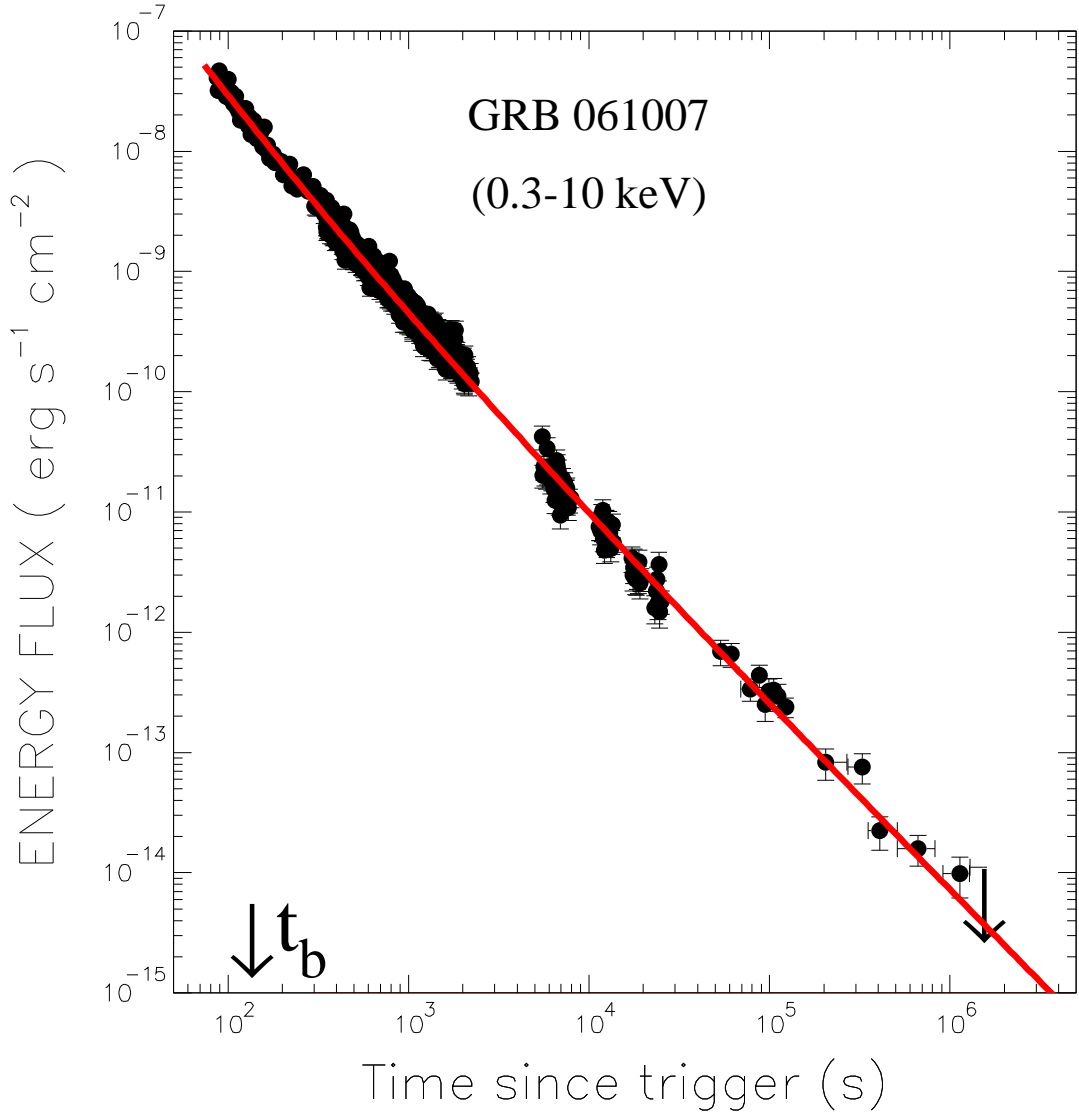,width=16.0cm,height=16.0cm}
\caption{The X-ray light-curve of GRB 061007 (Evans et al. 2007, 2009) and 
its CB model best fit, which have yielded an early break time hidden under 
the prompt emission phase and a post break late-time
$\alpha_X=1.61$ (Dado et al.~2008). The photon spectral index
$\Gamma_X=2.011\pm0.10$ reported in the Swift/XRT light-curve repositry 
satisfies within error the CB model relation $\alpha_X+1/2=\Gamma_X$.} 
\label{fig2}
\end{figure}

\newpage
\begin{figure}[]
\centering
\epsfig{file=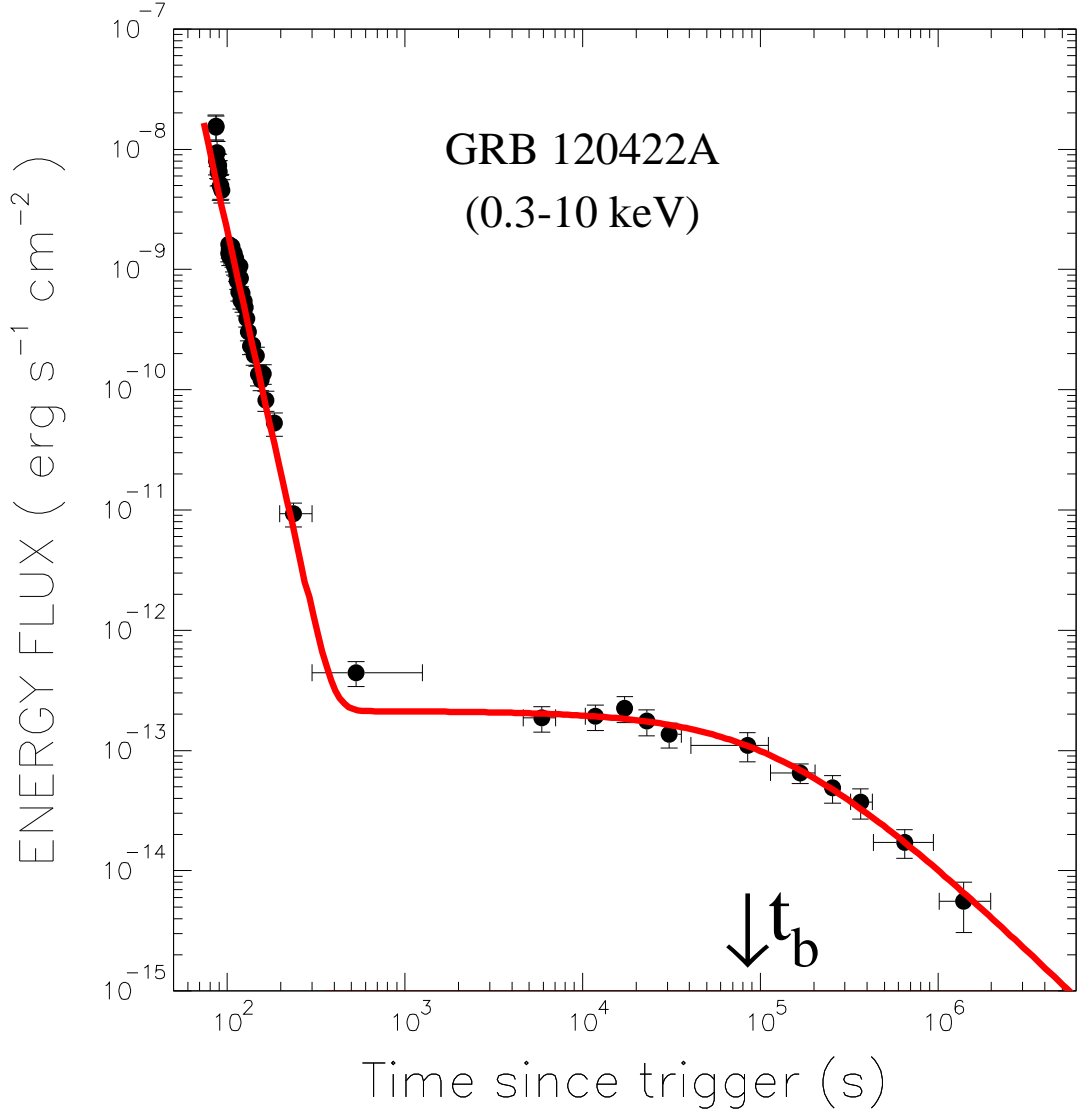,width=16.0cm,height=16.0cm}
\caption{The X-ray light-curve of GRB 120422A reported in 
the Swift/XRT light-curve
repository (Evans et al. 2007, 2009) and its CB model 
description assuming a constant
density ISM and the best fit values $\gamma(0)\, 
\theta=1.2$,~~$p_e/2=\beta_X=0.98\pm 0.04$ and 
a break time $t_b=84256$ s. The spectral index 
of the afterglow inferred from 
the Swift/XRT measurements in the PC mode is 
$\beta_X=1.03 (+0.38,-0.20)$.} 
\label{fig3}
\end{figure}

\newpage
\begin{figure}[]
\centering
\epsfig{file=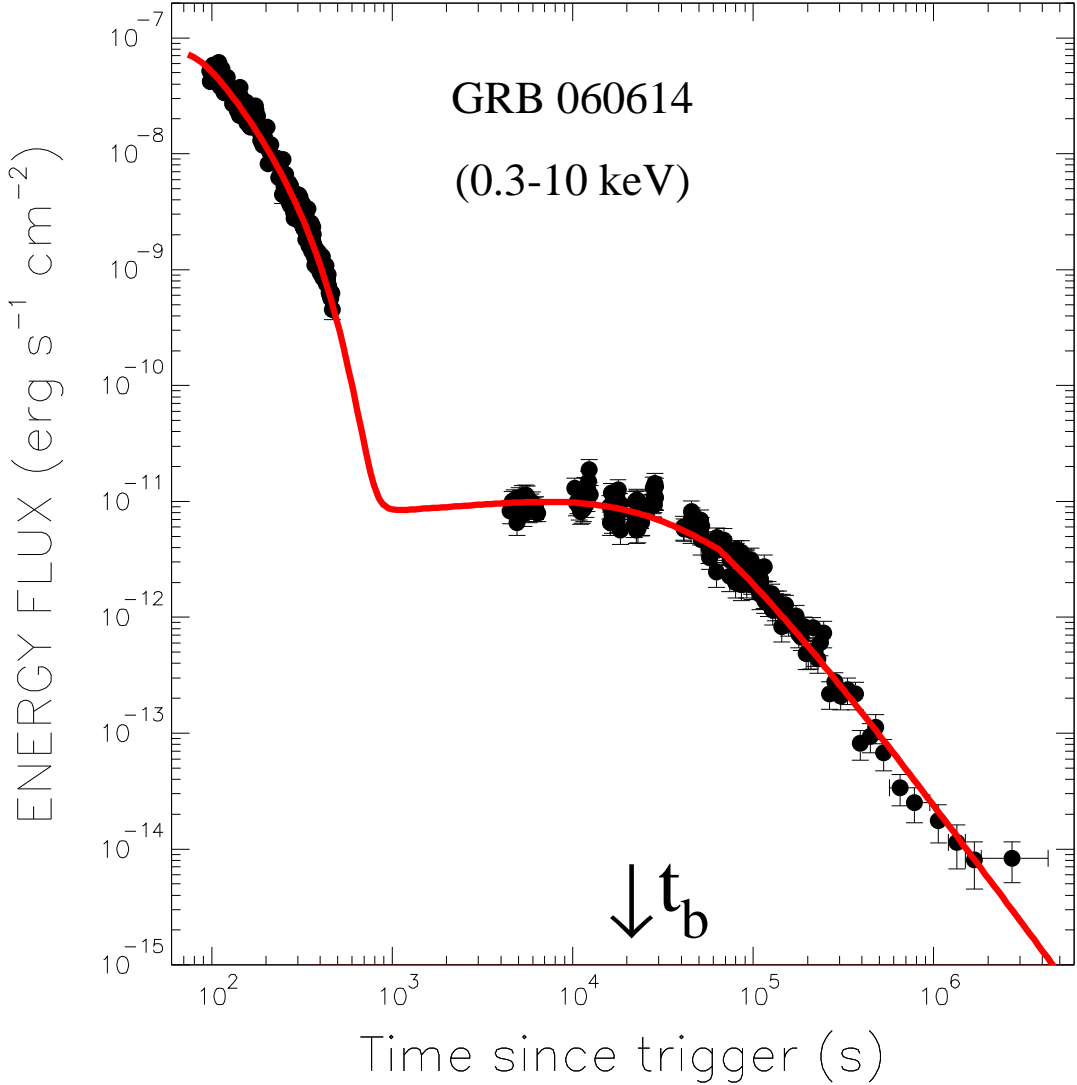,width=16.0cm,height=16.0cm}
\caption{
The light-curve of the 0.3-10 keV X-ray  
afterglow of the  short hard burst 
060614 and its CB model best fit 
(Dado et al. 2009b) assuming a late-time 
density profile  $ \propto r^{-2}$.
The best fit late-time decline yields $\alpha_X=1.98$,
which implies $\beta_X=0.98$  in agreement with the measured
spectral index $\beta_X=0.90 \pm 0.10$, which is
reported in the Swift/XRT light-curve repository (Evans et 
al. 2007, 2009).}
\label{fig4}
\end{figure}

\newpage
\begin{figure}[]
\centering
\epsfig{file=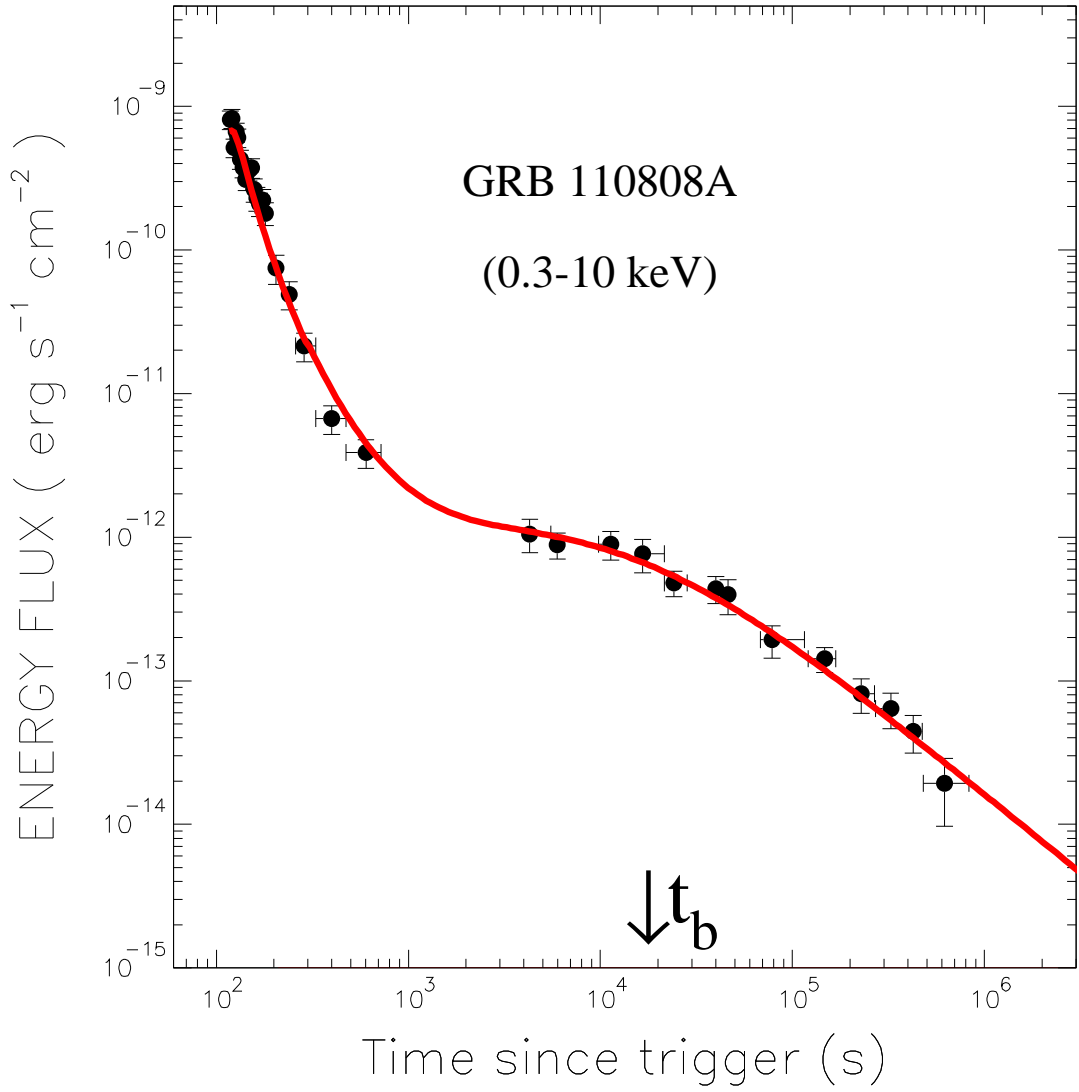,width=16.0cm,height=16.0cm}
\caption{
The light-curve of the 0.3-10 keV X-ray
afterglow of GRB 110808A 
(Evans et al. 2007, 2009) and its CB model
description assuming a shot gun configuration of CBs and the canonical 
value $\beta_X=1.1$. }
\label{fig5}
\end{figure}

\newpage
\begin{figure}[]
\centering
\epsfig{file=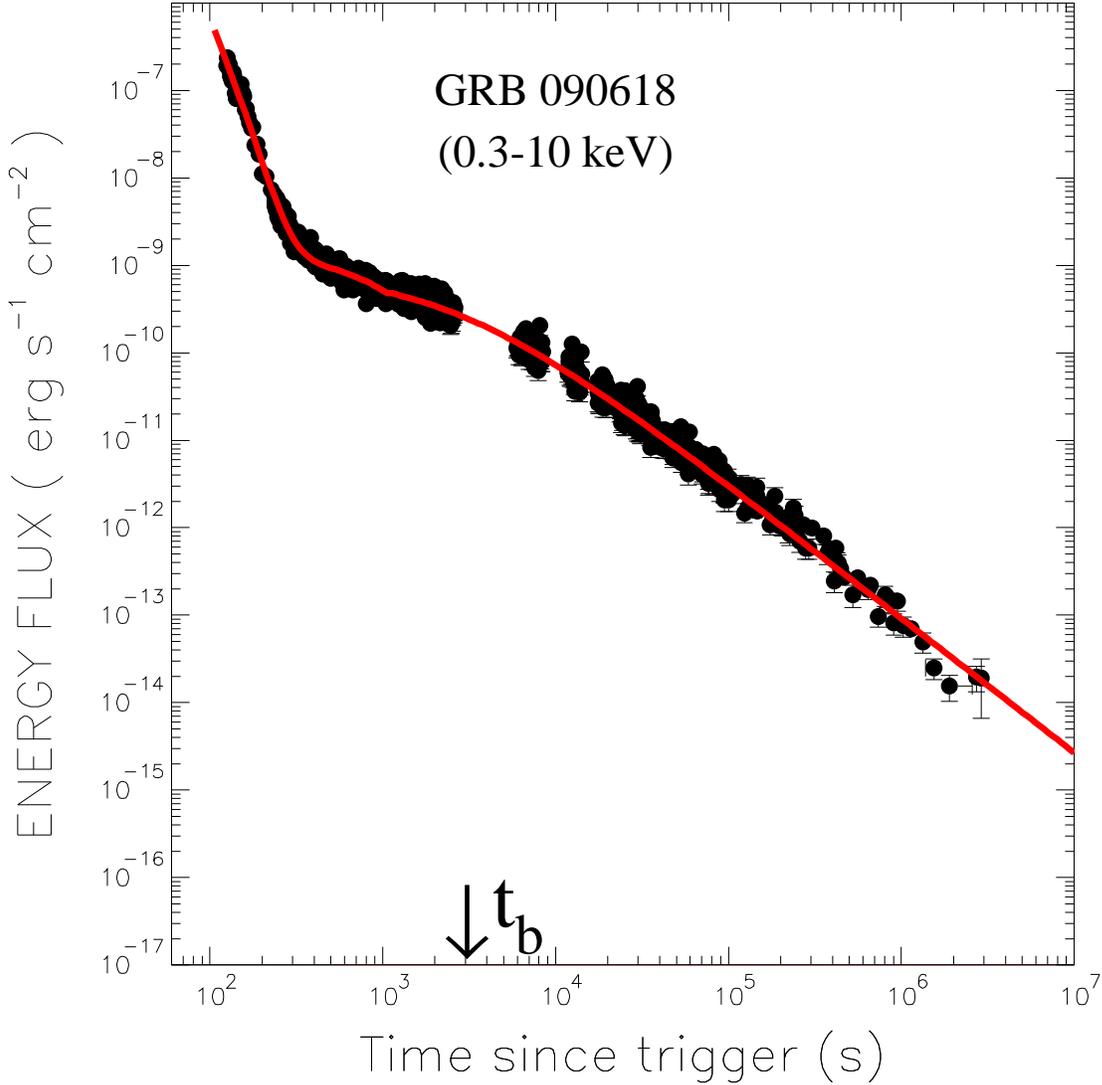,width=16.0cm,height=16.0cm} 
\caption{The X-ray light-curve of GRB 090618 reported in the Swift/XRT 
light-curves repository (Evans et al. 2007, 2009) and its CB model 
description assuming a constant density ISM. The best fit value 
$p_e=2.08\pm 0.09$ yields $\beta_X=p_e/2=1.04 \pm 0.03$, i.e., 
$\alpha_X=p_e/2+1/2=1.54\pm 0.03$ and $\Gamma_X=2.04$. The late-time 
spectral index inferred from the unabsorbed spectrum of the 0.3-10 keV 
X-ray afterglow measured with the Swift XRT (Evans et al. 2007, 2009) is 
$\beta_X=1.00 \pm 0.10$ (Cano et al., 2011).}

\label{fig6}
\end{figure}

\newpage
\begin{figure}[]
\centering
\epsfig{file=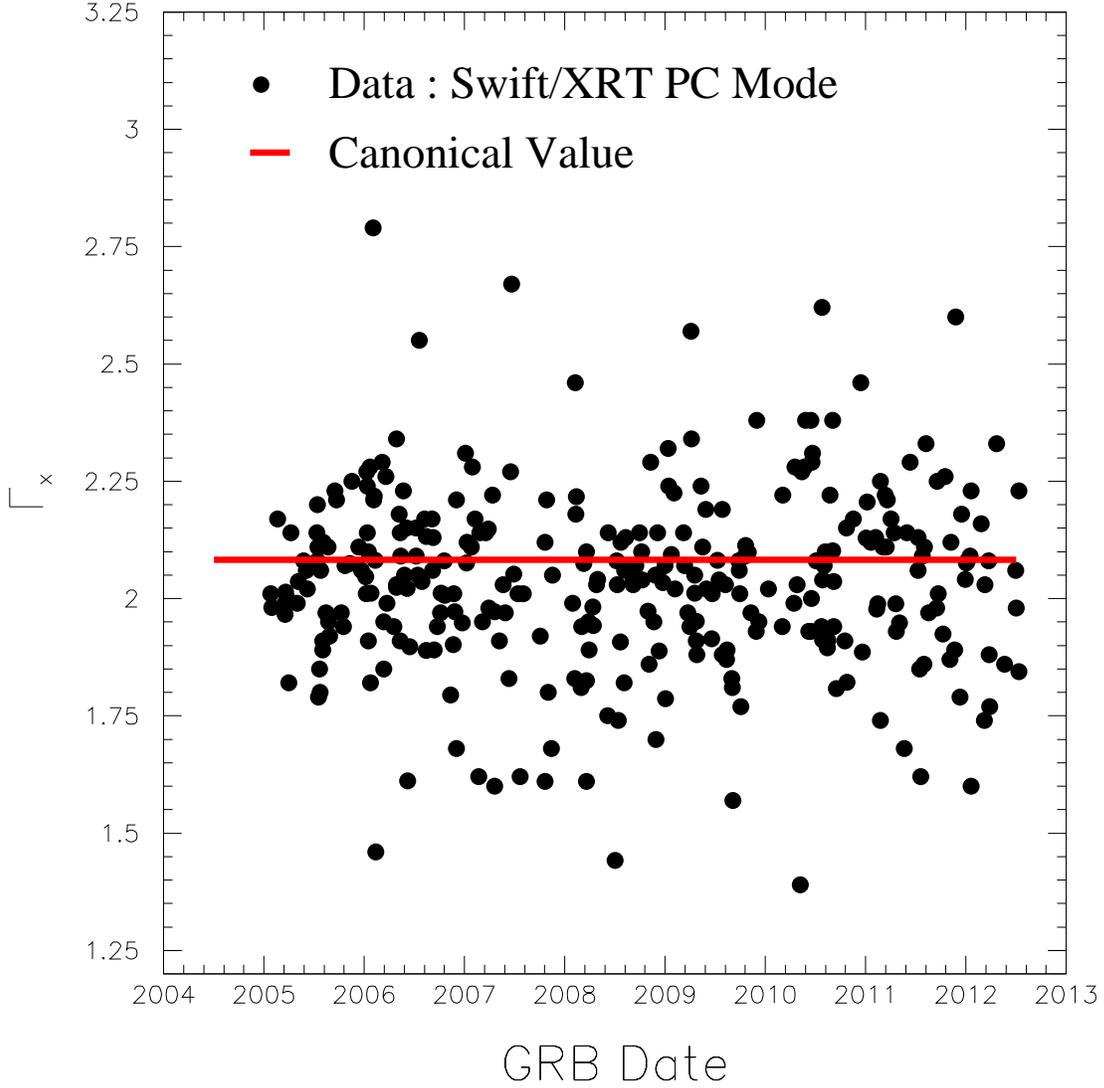,width=16.0cm,height=16.0cm}
\caption{   
The values of the  photon spectral index $\Gamma_X=\beta_X+1$ of the 
0.3-10  keV X-ray afterglow 
of the 315 Swift GRBs that were reported in the Swift/XRT light-curves 
repository (Evans et al. 2007, 2009) before August 1, 2012 
whose measured light-curve extended beyond 1 day and 
the estimated error in their photon spectral index is $<0.25$. 
The horizontal line represents the CB model canonical value 
$\Gamma_X=25/12=2.083$. 
}
\label{fig7}
\end{figure}

\newpage
\begin{figure}[]
\centering
\epsfig{file=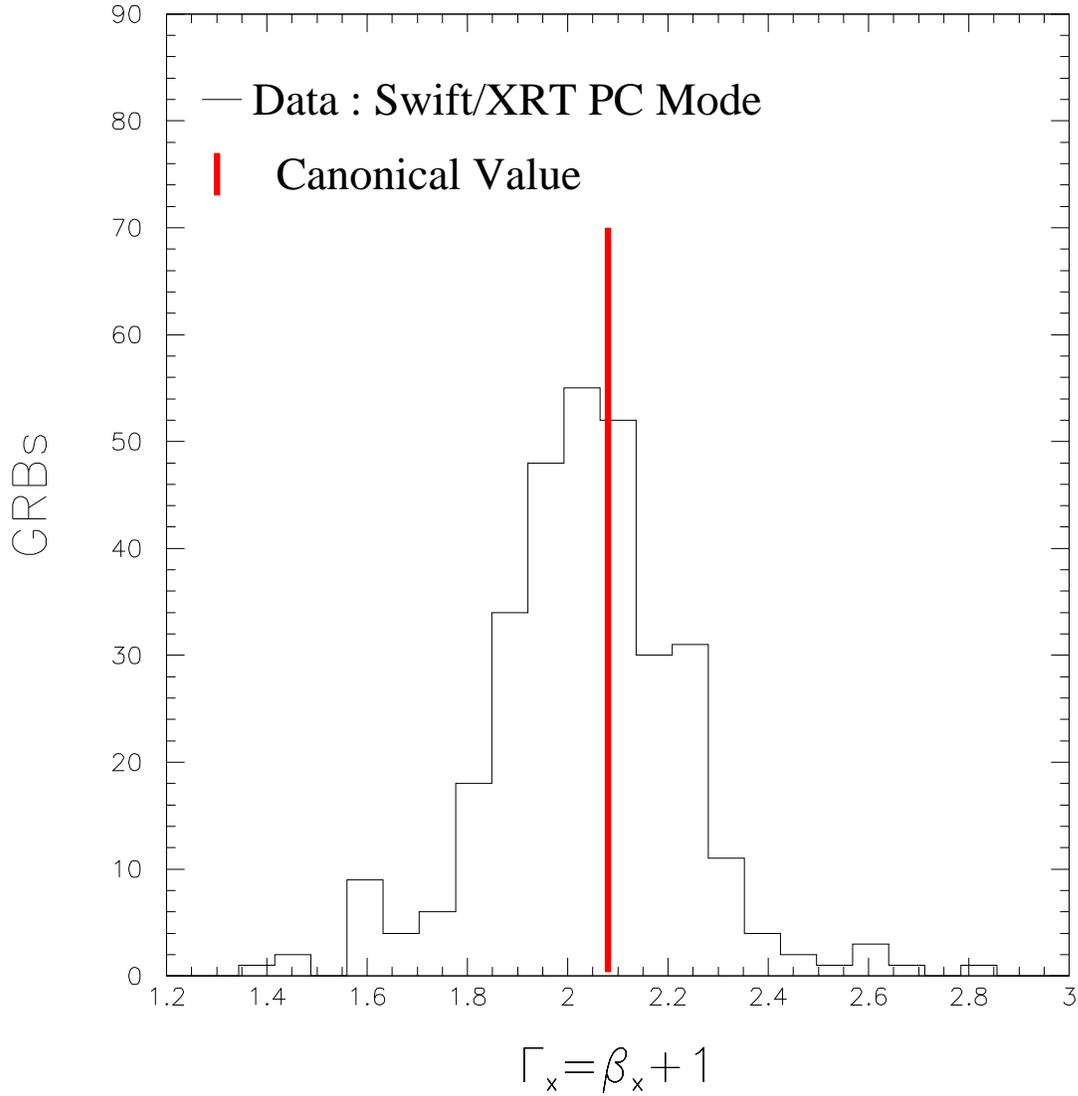,width=16.0cm,height=16.0cm}
\caption{   
The distribution of the {\it late-time}
photon spectral index $\Gamma_X$ measured with the Swift/XRT
in the PC mode for the 322 GRBs plotted in Fig. 1.
The vertical line is the CB Model  canonical value $\Gamma_X=25/12=2.08$.}
\label{fig8}
\end{figure}

\newpage
\begin{figure}[]
\centering
\epsfig{file=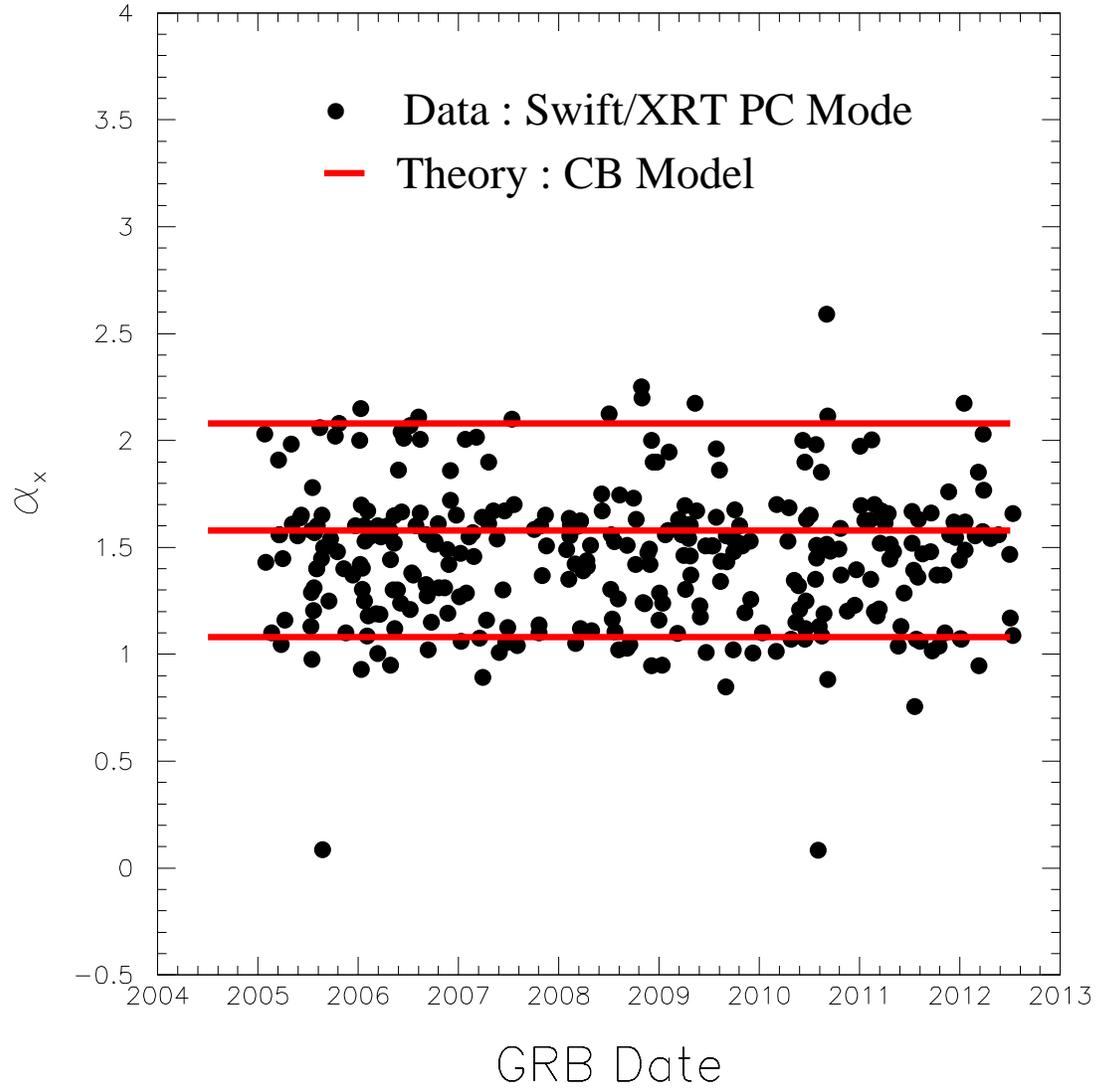,width=16.0cm,height=16.0cm}
\caption{
The values of the late-time temporal index $\alpha_X$ 
obtained  from the CB model best fits 
to the  the 0.3-10 keV X-ray afterglow
of the 315 Swift GRBs that were plotted in
Figure 7.
The horizontal lines represents the CB model canonical values
$\alpha_X\approx $1, 1.58 and 2.08.}
\label{fig9}
\end{figure}

\newpage
\begin{figure}[]
\centering
\epsfig{file=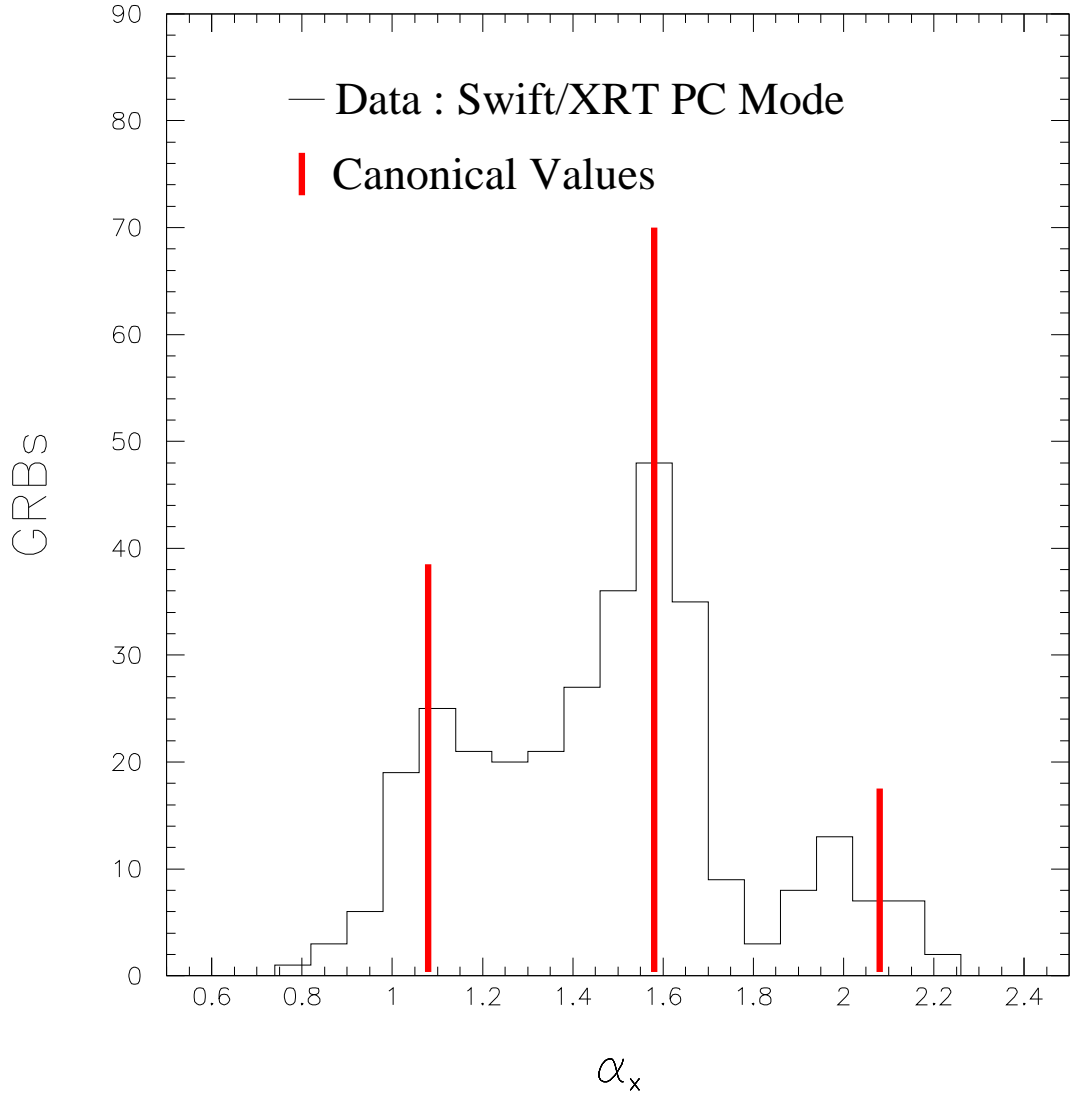,width=16.0cm,height=16.0cm}
\caption{
The distribution of the values of the late-time temporal index $\alpha_X$ 
obtained  from the CB model best fits
to the  the 0.3-10 keV X-ray afterglow     
of the 315 Swift GRBs that were plotted in  
Figure 9.
The vertical lines represents the CB model canonical values
$\alpha_X \approx $1, 1.58  and 2.08 .}
\label{fig10}
\end{figure}

\newpage
\begin{figure}[]
\centering
\epsfig{file=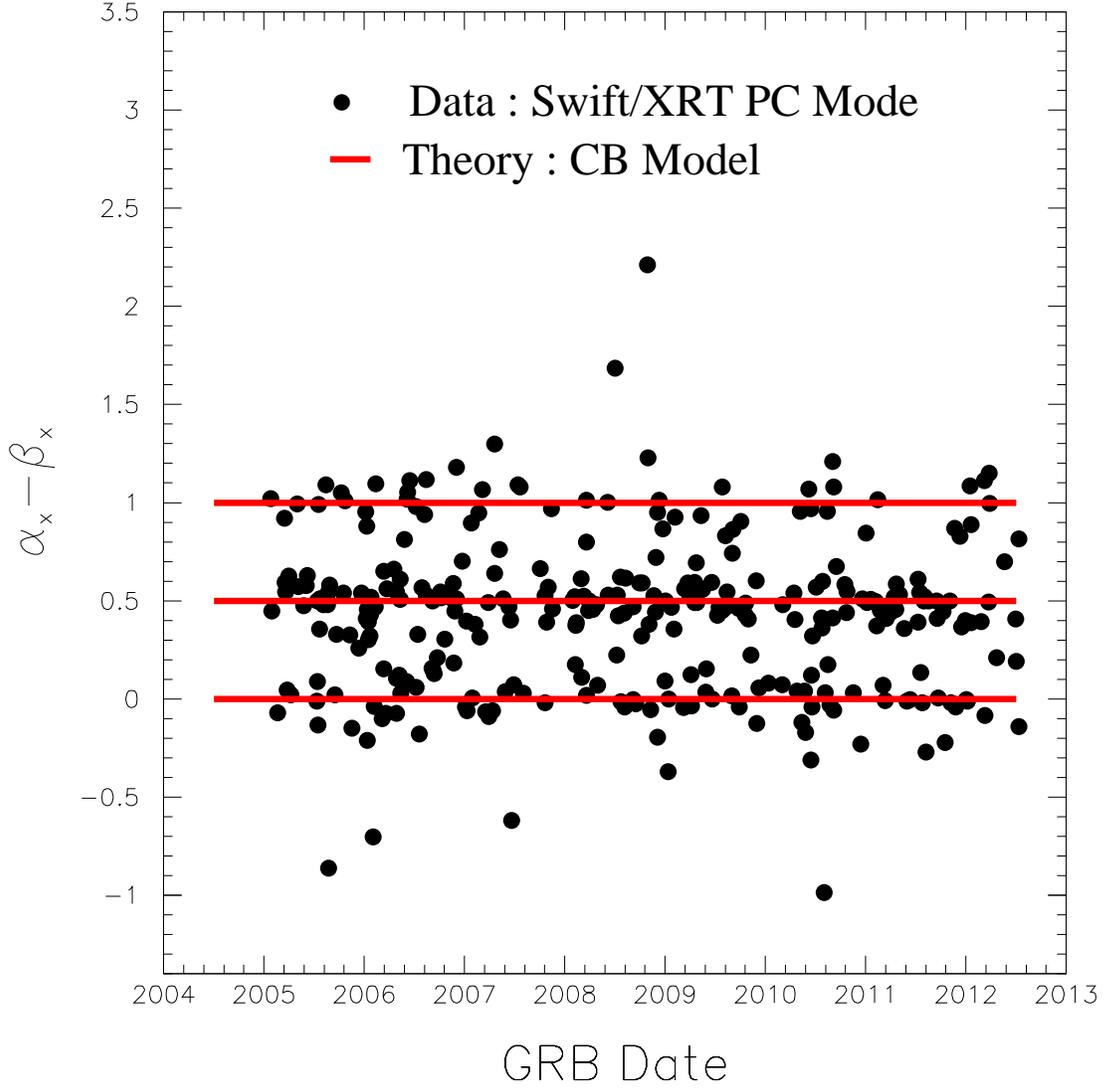,width=16.0cm,height=16.0cm}
\caption{   
The values of $\alpha_X-\beta_X$
for the late-time 0.3-10 keV X-ray afterglow of 315 Swift GRBs detected 
before August 1, 2012, that are plotted in Figures 7 and 9. 
The horizontal lines  represent the CB model expectation (see section 
2) $\alpha_X-\beta_X$=0,~~ 1/2 or  1 for error-free measurements.}
\label{fig11}
\end{figure}

\newpage
\begin{figure}[]
\centering
\epsfig{file=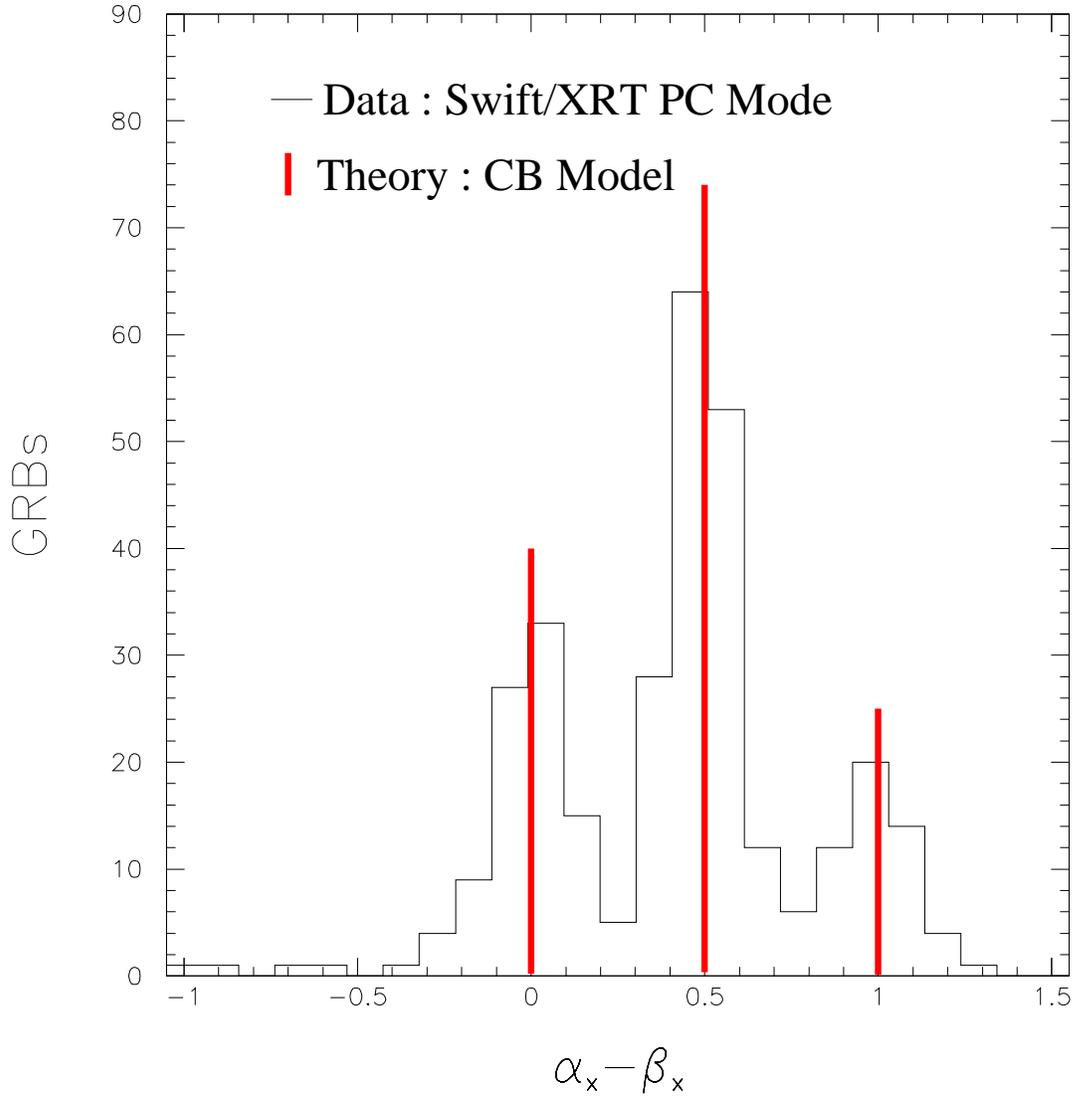,width=16.0cm,height=16.0cm}
\caption{   
The distribution of the values of $\alpha_X-\beta_X$
which  were plotted in Fig.~11  and the CB model expectation
$\alpha_X-\beta_X$=0,~~ 1/2 or  1 (vertical lines) for 
error free measurements.}
\label{fig12}
\end{figure}

\end{document}